\documentclass[twocolumn,showpacs,preprintnumbers,amsmath,amssymb]{revtex4}
\usepackage{graphicx}
\usepackage{epsfig}
\usepackage{dcolumn}
\usepackage{bm}

\begin{document}
\title{Spin control in semiconductor quantum wires}
\author{R.\ G.\ Nazmitdinov}
\affiliation{Departament de F{\'\i}sica,
Universitat de les Illes Balears, E-07122 Palma de Mallorca, Spain}
\affiliation{Bogoliubov Laboratory of Theoretical Physics,
Joint Institute for Nuclear Research, 141980 Dubna, Russia}
\author{K.\ N.\ Pichugin}
\affiliation{Kirensky Institute of Physics, Akademgorodok 50/38, 660036,
Krasnoyarsk, Russia}
\author{M.\ Val\'{\i}n-Rodr\'{\i}guez}
\affiliation{Conselleria d'Educaci\'o i Cultura, 
E-07004 Palma de Mallorca, Spain}
\date{\today}

\begin{abstract}
We show  that spin-flip rotation 
in a semiconductor quantum wire, caused by the Rashba and the Dresselhaus  
interactions (both of arbitrary strengths), can be suppressed by dint of an 
in-plane magnetic field. We found a new type of symmetry, 
which arises at a particular set of intensity and orientation of the magnetic field 
and explains this suppression.
Based on our findings, we propose a transport experiment
to measure the strengths of the Rashba and the Dresselhaus interactions.
\end{abstract}


\maketitle
Spin-polarized transport in semiconductor nanostructures is 
the main topic in spintronics due to great interests to both basic 
research and device application \cite{fab,aws}.
Spin-orbit interactions present in semiconductor structures 
provide a promising way to spin manipulation in bulk semiconductors \cite{6}, 
two-dimensional (2D) electron gases \cite{7}, and quantum dots \cite{8}.
However, these interactions cause decay of spin polarization \cite{dyak}, 
since the spin-orbit coupling  breaks the total spin symmetry.
The effect of spin relaxation produced by the interplay between the 
Dresselhaus \cite{dress} and Rashba \cite{rash} spin-orbit interactions (RDI) 
has been studied in a few publications (cf.\cite{av,sch,ross,loss,ohno,bern}). 
It was found by Schliemann {\em et al} \cite{sch} that  at zero magnetic field in 
2D  semiconductor nanostructures for equal strengths of the RDI  
there is an additional symmetry \cite{bern}. 
As a consequence, the orbital motion is decoupled 
from the spin evolution. If this resonant condition is not active, 
the spin dynamics is influenced by the different spin relaxation mechanisms related 
to orbital scattering processes.
In this paper we discuss another spin symmetry that arises at certain conditions
at nonzero magnetic field in the plane and arbitrary strengths of the spin-orbit terms in 
a quantum wire. In virtue of this symmetry the spin-flip rotation is suppressed
at arbitrary polarization of the injected electrons. By setting these conditions 'on' 
and 'off', the flow of a certain spin polarization through the device is either
allowed or destroyed, thus, defining a transistor-like action for the spin. 

We consider the conduction band of a 2D
semiconductor quantum well within the effective mass approximation. The wire
geometry is defined by a transversal potential '$V(y)$': 
${\cal H}_0=(p_x^2 + p_y^2)/{2m^*}+V(y)={\cal H}_y+p_x^2/{2m^*}$.
 The Dresselhaus interaction has, in general, 
a cubic dependence on the momentum of the carriers. For a narrow $[0,0,1]$ quantum well, 
it reduces to the 2D linear momentum dependent term 
${\cal H}_D={\beta}\left(p_x\sigma_x-p_y\sigma_y\right)/{\hbar}$ 
($\beta$ is the interaction strength). 
In the asymmetric quantum wells  the Bychkov-Rashba interaction 
has the form: 
${\cal H}_R={\alpha}\left(p_y\sigma_x-p_x\sigma_y\right)/{\hbar}$, where 
$\alpha$ is the corresponding strength. 
Our system Hamiltonian reads as 
\begin{equation}
\label{Ham}
{\cal H}={\cal H}_0+{\cal H}_D+{\cal H}_R+{\cal H}_Z\,,
\end{equation}
where we include the effect of the in-plane magnetic field by means of the
Zeeman interaction 
${\cal H}_Z=g^*\mu_BB\left(\cos\theta\sigma_x+\sin\theta\sigma_y\right)/2=
\varepsilon_z\left(\cos\theta\sigma_x+\sin\theta\sigma_y\right)/2$.
Here, $\theta$ represents the in-plane orientation of the magnetic field with the 
intensity $B$, $g^\ast$ is the effective gyromagnetic factor and 
$\mu_B$ is the Bohr's magneton. Note that none of the interactions 
break the translational invariance in the longitudinal coordinate. Therefore, 
the eigenstates are chosen to have a well-defined longitudinal momentum '$\hbar k$'
\begin{equation}
\label{Psi}
{\bf \Psi}_{nks}(\vec{r})= e^{ikx}\left(
\begin{array}{c}
\chi_{nks\uparrow}(y) \\
\chi_{nks\downarrow}(y)
\end{array}
\right)=:e^{ikx}\chi_{nks}(y),
\end{equation}
where $n$, $k$, $s$ stand for transversal, 
longitudinal and spin quantum numbers. As a result,
the Hamiltonian (\ref{Ham}) is transformed to the effective one for 
the transversal coordinate for a given value of '$k$'
\begin{eqnarray}
\label{Hk}
{\cal H}&=&{\cal H}_y+\frac{p_y}{\hbar}\left(\alpha\sigma_x-\beta\sigma_y\right)+\\
&+&\left[\left(\beta k+\frac{\varepsilon_z}{2}\cos\theta \right)\sigma_x -
\left(\alpha k-\frac{\varepsilon_z}{2}\sin\theta \right)\sigma_y\right].\nonumber
\end{eqnarray}
Two different spin-dependent terms can be distinguished within this Hamiltonian;
one is involving the transversal component of the momentum and  the  
other contains the effective Zeeman-like term including contributions 
from the RDI. If both terms are parallel in the spin space, a symmetry arises 
and the spin is totally decoupled from the orbital motion. In order to set 
this symmetry, it is required to fulfill the following condition 
\begin{equation}
\label{condition}
\frac{2\alpha k_0}{\varepsilon_z}\left[1-\left(\frac{\beta}{\alpha}\right)^2\right]=
\sin\theta+\frac{\beta}{\alpha}\cos\theta.
\end{equation}
Once Eq.(\ref{condition}) is fulfilled, the spin
operator ${\cal S}_{xy}=\alpha\sigma_x-\beta\sigma_y$ commutes with
the resulting Hamiltonian 
\begin{equation}
\label{hamiltonian}
{\cal H}={\cal H}_y+\left[\frac{p_y}{\hbar}+\frac{1}{\alpha}
(\beta k_0+\varepsilon_z \cos\theta/2)\right]\left(\alpha\sigma_x-\beta\sigma_y\right),
\end{equation}
i.e., $[{\cal H},{\cal S}_{xy}]=0$. 
Consequently, the spin symmetry is set up
for transversal eigenstates having longitudinal
momentum $k= k_0$. 
According to Eq.(\ref{condition}), this can be done by 
tuning a proper intensity $(\sim {\varepsilon_z})$  and an orientation of the 
applied in-plane magnetic field for  given strengths $\alpha$ and $\beta$.
It is noteworthy that this property is valid for any transversal potential
defining the wire geometry, since the symmetry arises from the relation between
the RDI and the Zeeman interaction in conjunction with the longitudinal
translational invariance. In addition, there is 
an extra degree of freedom, since the RDI strengths (for example, $\alpha$)
can be modified as well in order to fulfill  the condition \eqref{condition}.

In virtue of the spin symmetry, the spinorial part
of the eigenstates can be expressed as
\begin{equation}
\label{spinor}
\chi_s=\frac{1}{\sqrt{2}}\left(
\begin{array}{c}
1 \\
se^{-i\phi}
\end{array}
\right), \phi={\rm atan}(\beta/\alpha),
\end{equation}
where thereafter $s=\pm 1$.
These eigenspinors correspond to the in-plane orientation of the spin, where the 
particular orientation is determined by the ratio between the strengths of 
the both spin-orbit mechanisms. Note that in the Hamiltonian \eqref{hamiltonian} 
the spin-dependent term, linear in the transversal momentum '$p_y$', 
can be eliminated  by redefining the origin of the transversal momentum for 
each spin state. The only effect of this term on the energy spectrum is a constant 
shift that may be neglected by changing the energy origin.

At the condition (\ref{condition}) hold fixed, 
the spectrum of the system is composed of that corresponding to the spin-independent
orbital motion '${\cal H}_0$' ($\varepsilon^0_{nk}$), the constant shift and a 
contribution arising from a combination of the  RDI strengths and 
the Zeeman interaction
\begin{equation}
\label{spectrum}
\varepsilon_{nk_0s}=\varepsilon^0_{nk_0}-\frac{m^\ast}{2\hbar^2}r^2
+\frac{s}{\alpha}\left(\beta k_0+\frac{\varepsilon_z}{2}\cos\theta\right)r.
\end{equation}
Here we introduced the absolute magnitude of the RDI  strength vector
$r=\sqrt{\alpha^2+\beta^2}$. The above contribution represents a constant spin splitting for the 
eigenstates and its value depends on the longitudinal momentum, 
the RDI strengths, the particular orientation and the intensity of the applied 
magnetic field. At the preserved symmetry the eigenstates (\ref{Psi}) 
take the form:
\begin{equation}
{\bf \Phi}_{nk_0s}(\vec{r})=e^{-isy/l_{RDI}}e^{ik_0x}\psi^0_n(y)\chi_s\,,
\end{equation}
where $\psi^0_n(y)$ are the eigenstates of 
${\cal H}_y$. We have also defined the length $l_{RDI}=\hbar^2/(m^\ast r)$ 
giving the characteristic scale for the RDI strengths. 

At given Fermi energy $E_F$  
eigenstates \eqref{Psi} have a few real longitudinal momentum $k$.
Some of them have $k>0$ (propagating right), while the others have
$k<0$ (propagating left). One of those $k$ could satisfy the condition
\eqref{condition} by the adjusted magnetic field and, consequently, 
the corresponding spinor does not depend on coordinates. 
However, even a small mixing between the selected state and those that propagate
in the same direction but have different $k$ leads to the spin precession 
in the process of the propagation.
To suppress unwanted $k$-values one can adjust the Fermi energy 
(or the potential $V(y)$) to have only four real longitudinal momenta. Next, the
tuning of the intensity and the orientation of the magnetic field enables 
us to have two eigenstates (with $s=\pm$ in Eq.(\ref{spectrum})), 
propagating in the same direction, with the same energy and $k_0$. 
From Eqs.(\ref{condition}), (\ref{spectrum}) one obtains that
such a possibility can be realized, if 
the components of the magnetic field 
are proportional to the components of the RDI vector:
\begin{equation}
\label{Mf}
\frac{\varepsilon_z}{2}\cos\theta_0=
- k_0\beta, \quad
\frac{\varepsilon_z}{2}\sin\theta_0=
k_0\alpha .
\end{equation}
As a result, there is no a spin-flip process for any superposition of 
these eigenstates. 
Note that, in contrast to the spin-field transistor 
proposed in Ref.\cite{sch} whose effect is based on a particular input
spin polarization, in our case the spin-flip is absent for an arbitrary input  
spin polarization. Also, the magnetic field leads to a nonequivalence of
the electron transport from the left to the right and vise versa:
Eq. \eqref{Mf} is fulfilled for $-k_0$ at the condition $\theta_0\to\theta_0+\pi$.

To illuminate the found effect  in the electron transport 
we perform numerical calculations of the $S$-matrix in the
tight-binding model (cf \cite{Ferry}). To proceed we use a square lattice 
$n=n_x\hat{x}+n_y\hat{y}$ ($\hat{x}$ and $\hat{y}$ are vectors of a length 
$a_0$ in $x$ and $y$ directions, respectively; $a_0$ is the lattice constant, 
$n_x$ and $n_y$ are integers).
Within this approach our  Hamiltonian \eqref{Ham} has the following form
\begin{equation}
\label{tbHam}
\begin{array}{lcl}
{\cal H}_0&=&\sum\limits_{n,\sigma} \epsilon_{n\sigma} c^\dag_{n\sigma}c_{n\sigma}-
\sum\limits_{\langle nm\rangle,\sigma} t c^\dag_{n\sigma}c_{m\sigma}\\
{\cal H}_D&=&-\frac{\beta}{2a_0}\sum\limits_{n}
\Biggl\{
{\rm i}\left(c^\dag_{n\uparrow}c_{n+\hat{x}\downarrow}+
c^\dag_{n\downarrow}c_{n+\hat{x}\uparrow}\right)-\\ &&
\left(c^\dag_{n\uparrow}c_{n+\hat{y}\downarrow}-
c^\dag_{n\downarrow}c_{n+\hat{y}\uparrow}\right)
\Biggr\}+{\rm H.c.}\\
{\cal H}_R&=&-\frac{\alpha}{2a_0}\sum\limits_{n}
\Biggl\{
{\rm i}\left(c^\dag_{n\uparrow}c_{n+\hat{y}\downarrow}+
c^\dag_{n\downarrow}c_{n+\hat{y}\uparrow}\right)-\\ &&
\left(c^\dag_{n\uparrow}c_{n+\hat{x}\downarrow}-
c^\dag_{n\downarrow}c_{n+\hat{x}\uparrow}\right)
\Biggr\}+{\rm H.c.}\\
{\cal H}_Z&=&\frac{\varepsilon_z}{2}\sum\limits_{n}
c^\dag_{n\uparrow}c_{n\downarrow}
\left( \cos\theta-i\sin\theta \right)+{\rm H.c.}.
\end{array}
\end{equation}
Here $c^\dag_{n\sigma}$ creates an electron at site n with spin $\sigma$ and energy
$\epsilon_{n\sigma}=4t-V(n_ya_0)$, $t=\hbar^2/2m^\ast a_0^2$,
and $\langle nm\rangle$ stands for nearest neighbors sites $n$ and $m$. 
For the sake of illustration,
we choose for the wire potential a hard wall one: $V(y)=0$ for $0<y<W$ and  
$V(y)=\infty$ otherwise.  
\begin{figure}[t]
\begin{center}
\epsfig{file=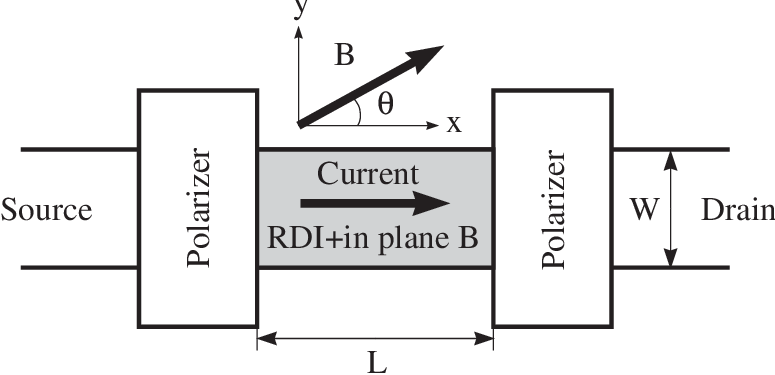, width=7cm} \caption{Sketch of the
2D wire device.} \label{fig1}
\end{center}
\end{figure}

Let us consider the system with a geometry shown in Fig.\ref{fig1}. 
It consists of a finite scattering area
with two lateral contacts. Each contact is
a narrow stripe with the width $W=20a_0$ and, for simplicity, no spin-orbit
couplings and no the magnetic field. The contacts are gated to
have two active channels (spin up and down) with a conductance $e^2/h$ in
each. Thus, the RDI and the in-plane magnetic field present only in
the scattering area of the length $L$ and the width $W$. The experiment may
consist of injecting a current $I$ through the left contact (source) to
the wire and measuring the voltage drop $V_R$ generated in the right contact
(drain). According to the Landauer-Buttiker formalism for linear response
(cf \cite{Al}), the ratio $V_R/I$ can be expressed by dint 
of the $S$-matrix elements $S_{m\sigma_2 n\sigma_1}$, where 
$n\sigma_1(m\sigma_2)$ denote the channels in the source (the drain).
In our approach the spin
resolved conductance between the source and the drain is determined as
$G_{\sigma_1\sigma_2}=e^2/h\int dE[-f^{\prime}(E-E_F)]
\sum_{nm}|S_{m\sigma_2 n\sigma_1}|^2$, where 
$f=1/[1+\exp{((E-E_F)/k_BT})]$. 
The conductance is calculated with  the energy dependent
$S$-matrix by direct solving the Schr\"odinger equation in a discretized space
according to the method suggested in Ref.\cite{ando}.
\begin{figure}[t]
\begin{center}
\epsfig{file=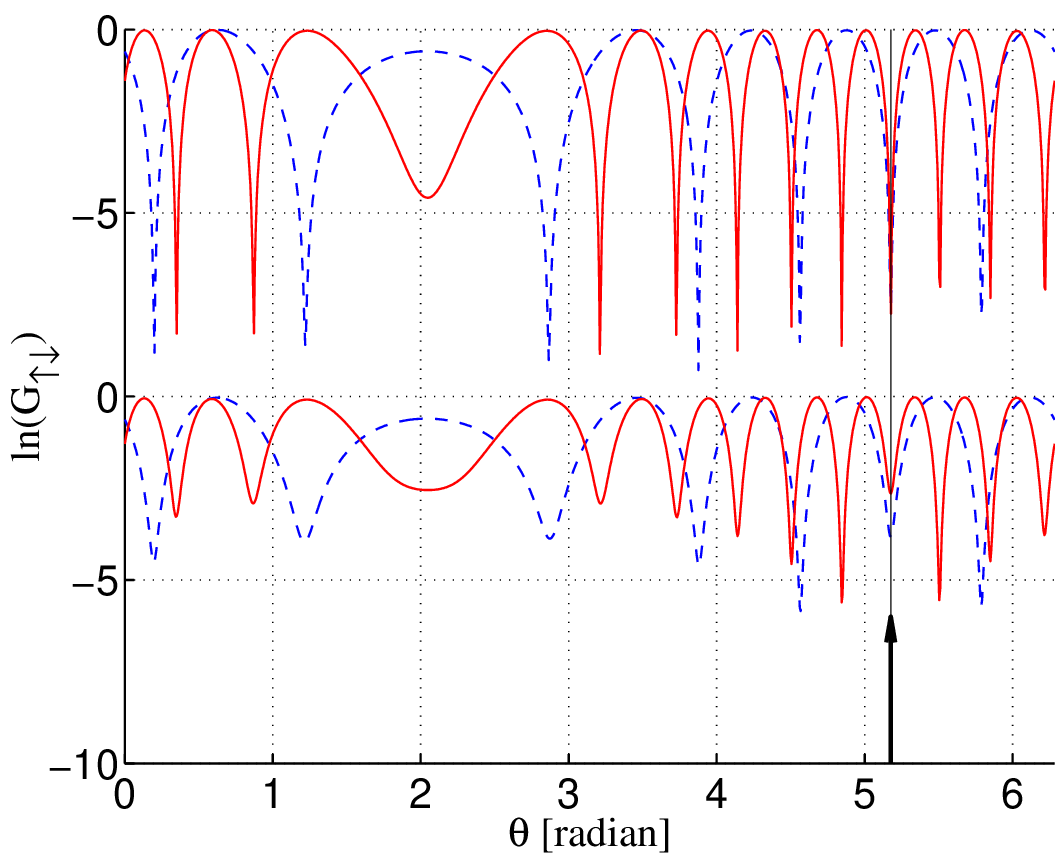, width=7cm}
\caption{(Colour online) The logarithm of the spin-flip conductance 
(in units of $e^2/h$) as a function 
of the magnetic field orientation $\theta$ 
(the intensity  $B=1.8$T)
for different sample lengths: $L$=31$W$ (solid line), $L$=17$W$ (dashed line).
Temperature is $T$=0K (top) and $T$=1K (bottom).
The incoming electrons are polarized along $z$-axis.
The set of parameters are typical for InAs: $E_F$=10 meV, $W$=45 nm, 
$\alpha$=20 meV nm, $\beta$=10 meV nm, $g^\ast$=-14.9,
$m^\ast$=0.023$m_e$. The arrow indicates the unique position 
of the angle $\theta_0$ for non-spin-flip conductance, 
independent on the sample length.}
\label{fig2}
\end{center}
\end{figure}

\begin{figure}[t]
\begin{center}
\epsfig{file=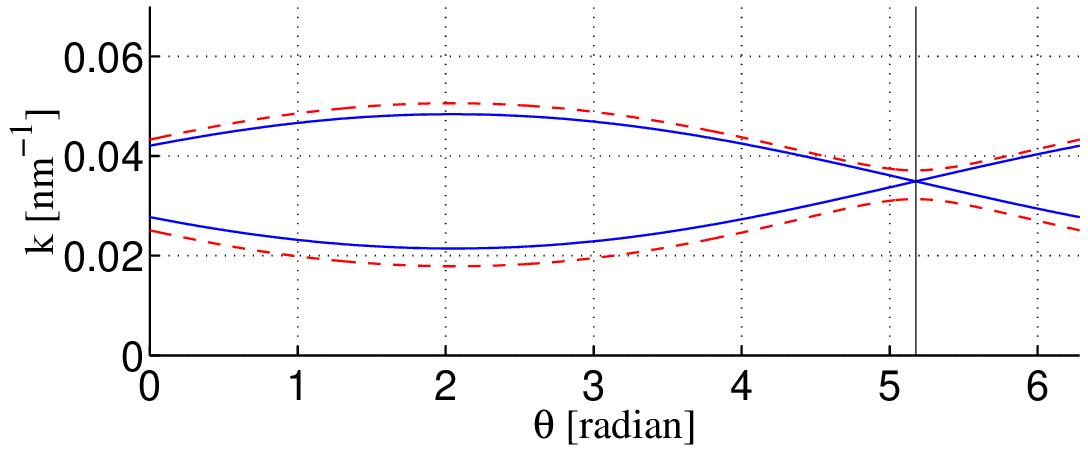, width=7cm}
\caption{(Color online) The longitudinal momentum $k$ 
as a function of the
orientation of the magnetic field (the angle $\theta$ 
) which intensity
is subject to Eq.\eqref{Mf} (solid line) and is slightly different (dashed
line). At the value $\theta_0=5.176$ (vertical line) the both positive $k$
coincide. The parameters are the same as for Fig.\protect\ref{fig2}.
}
\label{fig3}
\end{center}
\end{figure}

Note that the interfaces (the polarizers) between areas with and without the RDI
introduce some uncontrollable excitations of all modes inside the scattering
area. In particular, these excitations produce a  superposition 
(with coefficients $a_1$ and $a_2$) of two eigenfunctions 
$\chi_{1,2}(y)\exp(ik_{1,2}x)$ with different longitudinal momenta and, 
therefore, rotate the spin during a transport along the $x$-axis. 
Indeed, one has the following expectation values 
\begin{equation}
\left<S_x\right>= {\rm Re}{\xi},\quad \left<S_y\right> = {\rm Im}{\xi}, \\
\end{equation}
where ${\xi}=(a_1^*\chi^*_{1\uparrow}+a_2^*\chi^*_{2\uparrow})(a_1\chi_{1\downarrow}
+a_2 \chi_{2\downarrow})\times \exp(i(k_2-k_1)x)$.
Evidently, for equal longitudinal momenta, the expectation values are independent 
on the $x$ coordinate. The results  (see Fig.\ref{fig2}, top) 
manifest a single common minimum of the spin-flip conductance $(\sim 10^{-3} e^2/h)$
for one value of the magnetic field orientation but for different sample lengths, 
at a given intensity of the magnetic field and at zero temperature. 
At this value Eq.\eqref{Mf} holds, indeed. 
For another angles there is the mixing of wavefunctions
with different $k$ which leads to the electron spin rotation in the sample.
Fig.\ref{fig3} illuminates the dependence of longitudinal momenta $k(\theta)$ 
on the magnetic field orientation. At a particular value of the angle 
$\theta_0$ two wavenumbers coincide. However, the change of the magnetic field
intensity (the value of $\varepsilon_z$) leads  to avoided crossing of two 
curves $k(\theta)$. 

In real experiments the injected beam consists of electrons with different
energies due to, for example, a nonzero temperature. The temperature induces a
small mixture of spin-flip components and results in the increase of the
spin-flip conductance (see Fig.\ref{fig2}, bottom). However, it does not affect
the angle value at which the minima occur simultaneously in the two samples at
temperature $T=1$ K.

\begin{figure}[t]
\begin{center}
\epsfig{file=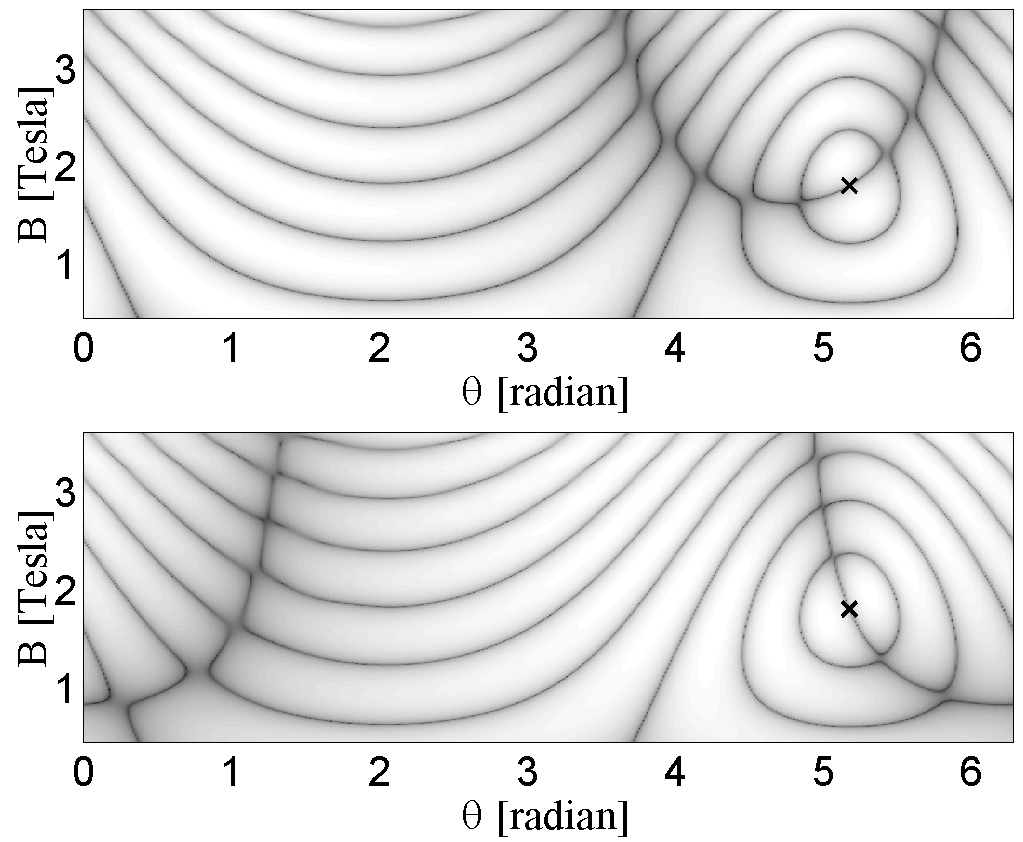, width=7cm}
\caption{The logarithm of the spin-flip conductance 
(in units of $e^2/h$) as a function of the intensity $B$ 
and orientation $\theta$ 
of the magnetic field for the sample length 
$L=31W$ and temperature $T$=0K (the darker is line  the lesser is the 
conductance).
The input polarization is along $x$ axis (top) and $y$ axis (bottom).
The minimum due to Eq.\eqref{Mf} is pointed by $\times$ mark.
The parameters are the same as for Fig.\protect\ref{fig2}.}
\label{fig4}
\end{center}
\end{figure}

Measurements of the RDI strengths is a subject of intensive experimental efforts
\cite{7,gan}.
For example, the ratio $\alpha/\beta$ could
determined with the aid of the analysis of photocurrents \cite{gan}.
Note that Eq.\eqref{Mf} enables us to determine the strengths 
of the Rashba and Dresselhaus interactions. 
We propose to use
a wire with a length determined by the condition $kL/(2\pi)\sim 5$.
According to our analysis (see Fig.\ref{fig4}), such a system 
produces a few,  well resolved spin-flip conductance 
minima. This condition helps also to diminish the effect of evanescent modes.
The Fermi energy and the transversal potential $V(y)$ should be taken to 
support only four
propagating modes (two with a positive $k$, going to the right 
and two with a negative $k$, going to the left).
The measure of the spin-flip conductance provides  a set of spin-flip minima
at different intensities and different orientations of the magnetic field, at
a fixed input polarization. Taking another polarization and repeating the same 
measurement, one obtains a different pattern for the location of the minima. 
As an example, we calculate the spin-flip conductance for two
input polarizations -- along $x$- and $y$- axis (see Fig.\ref{fig4}).
One obtains the required minimum which is subject to Eq.\eqref{Mf} 
at the same angle and the same intensity in different setups, since the effect 
is independent on the polarization. One might repeat measurements for 
different sample lengths, since the minimum position is independent 
on the length too. To  diminish the effect
of multiple reflection from the polarizers we suggest 
to use the same polarization direction in the both polarizers.

In conclusion, we found the condition (Eq.(\ref{condition})) 
to decouple the spin and the orbital motion of electrons in a quantum wire 
with the in-plane magnetic field and arbitrary Rashba and Dresselhaus strengths. 
At this condition there is the spin symmetry in  an arbitrary  transversal 
potential defining the wire geometry. Furthermore,
at the specific condition (\ref{Mf}) the magnetic field 
cancels the RDI for the electron momentum $k_0$. 
As a result, during the electron transport through the wire 
the spin-flip rotation is absent
for any chosen polarization. We propose the
experiment to measure the strengths of the Rashba and
the Dresselhaus interaction by finding the minimum of the spin-flip conductance, 
which should occur at the condition (\ref{Mf}).

\section*{Acknowledgements}
This work is partly supported by
Grant No. FIS2008-00781/FIS (Spain) and
RFBR Grant No. 08-02-00118 (Russia).

\end{document}